\documentclass[10pt,aps,prd,nofootinbib,nobibnotes,longbibliography,floatfix,twocolumn]{revtex4-2}

\usepackage{bm}
\usepackage{mathtools,
amsmath,
amssymb,
amsfonts,
mathrsfs,
chngcntr,
multirow}

\let\cc\corresponds
\let\corresponds\relax
\usepackage{mathabx}
\let\corresponds\cc

\usepackage[utf8]{inputenc}
\usepackage[T1]{fontenc}

\usepackage[dvipsnames]{xcolor}
\usepackage[unicode]{hyperref}
\hypersetup{colorlinks=true, citecolor=MidnightBlue,
            linkcolor=MidnightBlue, urlcolor=MidnightBlue, linktocpage=true}
\usepackage[normalem]{ulem}

\DeclareMathAlphabet{\mathpzc}{OT1}{pzc}{m}{it}
\definecolor{darkgreen}{rgb}{0.0, 0.6, 0.0}

\newcommand{\note}[1]{\text{\scshape\tiny{#1}}}

\newcommand{\ii}{\mathrm{i}}
\newcommand{\dd}{\mathrm{d}}

\newcommand{\F}{\mathcal{F}}

\newcommand{\T}{\mathcal{T}}

\newcommand{\de}{\delta}

\newcommand{\la}{\lambda}

\newcommand{\sg}{\sigma}

\newcommand{\om}{\omega}

\begin{document}
\title{Quasi-normal modes ratios as agnostic test of general relativity}

\author{Nicola Franchini}
\email{nicola.franchini@tecnico.ulisboa.pt}
\affiliation{CENTRA, Departamento de F\'{\i}sica, Instituto Superior T\'ecnico -- IST, Universidade de Lisboa -- UL, Avenida Rovisco Pais 1, 1049-001 Lisboa, Portugal}

\begin{abstract}
In this letter, we provide a novel test of general relativity based on ringdown analysis. The test is performed on agnostic models, where the postmerger signal is fitted with a superposition of damped sinusoids. If at least two modes are detected, one has to compute the ratio of the frequencies and of the damping times and compare them against the predictions of general relativity. By considering ratios, the dependency on the black hole's mass is scaled away. Most notably, we find that the ratios vary very little with the spin, the real part depends mostly on the angular momentum of the mode $\ell$ and the imaginary part depends mostly on the overtone number $n$: different combinations create specific mode islands. We provide a qualitative explanation of these islands through a semi-analytical argument. We discuss the application of the method to future detectors. Finally, we show that ratios in alternative theories of gravity or between different field content drastically differ from those of general relativity. 
\end{abstract}

\maketitle

\noindent \textbf{\textit{Introduction:}}~Perturbation theory is the framework to describe the last stage of black hole (BH) binary mergers~\cite{Kokkotas:1999bd,Nollert:1999ji,Berti:2009kk,Konoplya:2011qq,Franchini:2023eda,Berti:2025hly}. This gravitational wave emission is referred to as ringdown, one of the cleanest ways to test BHs in general relativity (GR): analogously to atom spectroscopy, the aim of BH spectroscopy is to reconstruct the properties of BHs through their characteristic frequencies of oscillation. The perturbations of BHs are described by an infinitely discrete set of complex quasi-normal modes (QNMs), whose value depends on the mass of the BH only by a multiplicative factor and by its spin~\cite{Leaver:1985ax}.

Isolating the ringdown from the full binary BH signal is a long established test since the advent of gravitational waves astronomy~\cite{Echeverria:1989hg,Finn:1992wt}. The idea is to cut and keep the signal only after a time at which the nonlinearities have disappeared and fit it with a superposition of damped sinusoids. Different assumptions can be made on the model depending on the amount of information from GR one wants to use. Here, we restrict ourselves to a fit performed over 4 free parameters for each mode: its complex frequency and its complex amplitude. Further models that reduce the number of free parameters including notions from GR will not be discussed here~\cite{Carullo:2019flw,Isi:2021iql,Ghosh:2021mrv,LIGOScientific:2021sio,Gennari:2023gmx}.

Ringdown tests of GR are possible only when at least two modes are detected, to remove the mass/spin degeneracy in Kerr QNMs. Usually, these tests are performed by assuming Kerr predictions for the QNMs by measuring the final mass and final spin of the BH remnant. They generally fall in two categories: consistency test between ringdown against full inspiral-merger-ringdown (IMR) analysis; and tests adding of one or more free parameters to the QNMs to be constrained by the analysis~\cite{LIGOScientific:2021sio}. Both tests are well established in gravitational waves data analysis, and despite their solidity they might suffer from some subtleties.

All IMR consistency tests heavily rely on the accuracy of the results of IMR analysis, which can be distorted by systematic biases, unmodeled physics, non-GR effects~\cite{Gupta:2024gun,Dhani:2024jja,Kapil:2024zdn}. The advantage of a spectroscopy analysis independent from IMR is to get rid of all the effects which might modify the inspiral and the merger.
For the other test, frequency deviations are a good tool to put agnostic constraints on QNMs, but they have two potential issues. The set of modes assumed to be present in the model is arbitrarily based on GR considerations [{\it e.g.}, the $(\ell,m,n)=(2,2,0)$ mode is the most excited in GR] and furthermore assumes that deviations are on top of tensor modes, while we know that mode contamination between different fields might happen~\cite{Pani:2009wy,Yunes:2009hc,Pani:2011gy,Maselli:2015tta,Pierini:2021jxd,Pierini:2022eim,Wagle:2021tam,Srivastava:2021imr,Crescimbeni:2024sam,Lestingi:2025jyb}. 

In this letter, we propose a novel ringdown test purely based on damped sinusoids analysis. The idea is to perform a 2-modes analysis, identify the dominant mode and compute the ratios of both frequencies and damping times. By construction, these ratios do not depend on the mass and they mildly depend on the spin, as it was already observed for a few cases~\cite{Isi:2021iql,Ota:2021ypb}. Here, we provide a more comprehensive study case, surveying a vaste range of mode combinations and see that the trend is respected, at least for BHs far from extremality.

\begin{figure}
\centering
\includegraphics[width=\columnwidth]{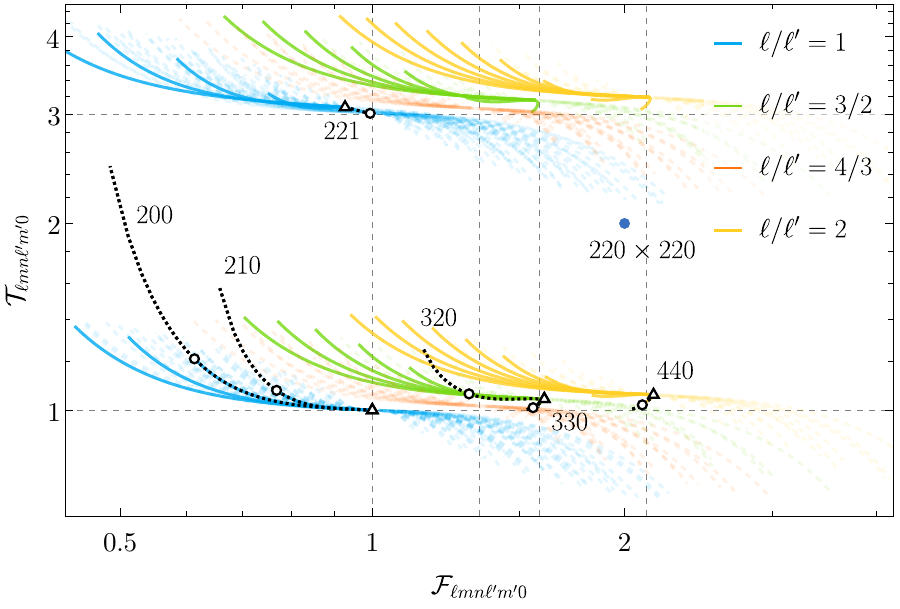}
\caption{Frequency ratio $\F_{j j'}$ versus damping time ratio $\T_{j j'}$ for various combination of $j, j'$. The black dotted lines correspond to $j'=(2,2,0)$ and selected values of $j=(2,2,1),(2,1,0),(2,0,0),(3,3,0),(3,2,0),(4,4,0)$, for values of the spin $a=[0,0.99]$; an open triangle marker is drawn at $a=0$, and a circular one at $a=0.9$. The solid lines represent correspond to $j'=(2,2,0)$ and $j=(\ell,m,n)$, with $\ell=[2,4]$, $m=[-\ell,\ell]$, $n=[0,1]$ for values of the spin $a=[0,0.9]$; different colours correspond to mode combinations of same $\ell/\ell'$. The cyan circle highlights the ratio between the quadratic mode $j=(2,2,0)\times (2,2,0)$ and the fundamental mode $j'=(2,2,0)$. The dashed-opaque lines correspond to $j=(\ell, m, n)$ and $j'=(\ell',m',0)$, with $\ell=[\ell',4]$, $m,n$ varying as before and $\ell'= [2,4]$, $m'=[-\ell',\ell']$, for values of the spin $a=[0,0.9]$. The gridlines are obtained from WKB estimates, see equations~\eqref{eq:F_WKB}--\eqref{eq:T_WKB}. \label{Fig:lmn-220}
}
\end{figure}

\noindent \textbf{\textit{The test:}}~We consider pair of modes $(\om_j, \om_{j'}$, each of which can be decomposed in real and imaginary part
$\om_j = 2\pi f_j - \ii/\tau_j$. For a Kerr BH, the QNMs are labeled by their angular momentum $\ell$, azimuthal number $m$ and overtone number $n$, such that one can make the identification $\om_j = \om_{\ell m n}$. Moreover, at second-order in perturbation theory quadratic QNMs are excited, with frequency $\om_j = \om_{\ell_1m_1n_2 \times \ell_2m_2n_2}$ from the combination of linear modes $\om_{\ell_1m_1n_1}$ and $\om_{\ell_2m_2n_2}$~\cite{Gleiser:1995gx,Cheung:2022rbm,Mitman:2022qdl}.
Let us define the ratios of the frequencies and damping times of two different modes as
\begin{equation}
\F_{j j'} = \frac{f_{j}}{f_{j'}} \qquad \T_{j j'} = \frac{\tau_{j'}}{\tau_{j}} \,.
\end{equation}
The ratios $\F_{j j'}$ and  $\T_{j j'}$ depend only on the spin, as the mass dependency cancels by definition. It is less immediate to infer the behavior of these two quantities with respect to $\ell m n$ and $\ell' m' n'$. Let us analyse it with steps, each of them reported in the summary figure~\ref{Fig:lmn-220}. First, we highlight the ratios that are mostly relevant for current BH spectroscopy. We assume that the $(2,2,0)$ mode is detected along with another mode. We fix $j'$ to this value and we vary $j$ among the combinations of $\ell,m,n$ which dominate the postmerger of a BH binary. For quasicircular, non-precessing binaries the dominant modes are $j=(2,2,1),(2,1,0),(2,0,0),(3,3,0),(3,2,0),(4,4,0)$ plus the quadratic $j=(2,2,0)\times(2,2,0)$~\cite{London:2014cma,Cheung:2023vki,Pitte:2023ltw,Capuano:2025kkl}. We report these ratio combination in figure~\ref{Fig:lmn-220} with black-dotted lines, for spin $a$ in $[0,0.99]$, denoting $a=0$ with a triangle and $a=0.9$ with a circle.

A few considerations relevant for BH spectroscopy are drawn:
\begin{itemize}
\item The lines corresponding to the ratios between the fundamental mode and the first overtone $(2,2,1)$, the $(3,3,0)$ mode and the $(4,4,0)$ mode are extremely localized around the points $(1,3)$, $(1.5,1)$, $(2.1,1)$, respectively, independently of the spin; 
\item For values of the spin $a\lesssim0.9$, the damping times ratios remain approximately constant: $\T_{0 0} \sim 1 $ and  $\T_{1 0} \sim 3$.
\item Varying the value of the angular momentum $\ell$ increases the ratio of the real parts $\F_{\ell 2}$ proportionally to $\sim \ell/2$, and the effect of the spin is just pushing $\F$ more to the left of these values.
\item The contribution from the quadratic mode $(2,2,0)\times (2,2,0)$ lies on the point (2,2) in the plane $\F - \T$.
\end{itemize}

The modes considered above form a subset of possibly excited modes in the ringdown. Features like mass ratio, line of sight, precession and eccentricity might excite different modes. Hence, to be more agnostic, and to verify the trend seen from the small batch of modes, we plot in figure~\ref{Fig:lmn-220} the ratios for all the combinations $j'=(2,2,0)$ and $j=(\ell, m, n)$ with $\ell = [2,4]$, $m=[-\ell, \ell]$, $n=[0,1]$, varying the spin between $0$ and $0.9$. The values are drawn with solid coloured lines. Moreover, in a most general setup, we relax the assumption that the $(2,2,0)$ mode is identified among the two modes, and we just assume that the mode $j'$ is identified as the least damped mode. This assumption fixes the overtone number $n'=0$ but $\ell'$ and $m'$ are varied in the same ranges as above: this results in the coloured-opaque dashed lines. In all the cases the colours are varied according to the ratio $\ell/\ell'$.

The considerations drawn from the full plot are following:
\begin{itemize}
\item The ratio of the frequencies $\F_{j j'}$ is mostly affected by the spin of the BH and by the angular momenta of the two modes $\ell$ and $\ell'$
\item The ratio of the damping times $\T_{j j'}$ is mostly affected by the overtone numbers of the two modes $n$ and $n'$, and by the spin of the BH only when it is quite large ($a >\sim 0.9$).
\item Given two combination $\ell_1, \ell_1'$ and $\ell_2, \ell_2'$ with $\ell_2/\ell_2' > \ell_1/\ell_1'$, we observe that $\F_{\ell_1 \ell_1'} < \F_{\ell_2 \ell_2'}$ and $\T_{\ell_1 \ell_1'} \simeq \T_{\ell_2 \ell_2'}$, for any value of the spin.
\item There appear to be clear overtone islands along the lines $\T_{n n'} \simeq \frac{2n+1}{2n'+1}$.
\end{itemize}

We remark that these results include also modes with limited relevance in BH spectroscopy, such as the retrograde modes (those with $m<0$) and the overtones of the higher modes $(3,m,1)$ and $(4,m,1)$. To avoid cluttering in the plot, we chose a maximum for $\ell=4$ and for $n=1$, but we have verified that our conclusions remain valid up to $\ell=7$ and $n=3$.

An important consideration is the ordering degeneracy in the test: for any pair of frequencies $j$ and $j'$, both the $(j,j')$ and $(j',j)$ orderings contribute to the plot as one the inverse of the other. There are several ways to break this degeneracy, through selection ordering rules:
\begin{itemize}
	\item Hierarchical ordering: as used in the previous paragraphs, one can keep only the ratios by selecting the most dominant frequency for $j'$;
	\item Frequency ordering: selects $j'$ as the mode with largest $f$;
	\item Damping time ordering: selects $j'$ as the mode with largest $\tau$;
	\item Inversion rule: exploits the relation $\F_{j j'} = 1/\F_{j' j}$ and $\T_{j j'} = 1/\T_{j' j}$ selecting only couples satisfying $\T_{j j'} \leq \F_{j j'}$ and replacing non-conforming pairs with the inverse with $1/\F_{j j'}$ and $1/\T_{j j'}$.
\end{itemize}

\noindent \textbf{\textit{Explanation of mode islands:}}~ We can find a qualitative explanation of the mode islands by QNM theory arguments. The Wentzel-Kramers-Brillouin (WKB) method provides a semianalytical way to express the eigenvalues of the equation $d^2 \Psi/d x^2 + (\om^2 - V) \Psi = 0$, provided that the effective potential $V$ has only one maximum, two turning points, and QNM boundary conditions~\cite{Bender:1999box,Schutz:1985km,Iyer:1986np,Iyer:1986nq}. At first order, the WKB formula yields
\begin{equation}
	\om^2 = V_0 - \ii \left(n+\frac{1}{2}\right) \sqrt{-2 V_2} 
\end{equation}
where $V_j = \dd^j V/ \dd x^j |_{x = x_M}$ is the $j$-th derivative of the potential $V$ with respect to the coordinate $x$ evaluated at the peak of the potential $x_M$. For a Schwarzschild BH, the linear perturbation equation is the Regge-Wheeler equation~\cite{Regge:1957td}, which can be recast in this form, with $V(r) = f(r)[\ell(\ell + 1)/r^2 - 3/r^3 ]$ and $\dd x = \dd r/f(r)$, being $f(r) = 1 - 1/r$ and $r$ is the areal radius in units of the BH's horizon radius $r_\note{h}=1$. In the eikonal limit $\ell \to \infty$, one has $r_M = 3/2$, from which $V_0 = 4(\ell^2 + \ell -2)/27$ and $V_2 = -32(3\ell^2+3\ell-8)/2187$. With some simple algebra, one finds that the WKB formula predicts
\begin{equation}
	\om^\note{R} \simeq - \frac{n^2 + n -2 +2\ell^2 + \ell}{3\sqrt{3}\ell}\,, \quad \om^\note{I} \simeq -\frac{2n+1}{3\sqrt{3}} \,,
\end{equation}
from which we obtain
\begin{align}
	\F_{\ell n \ell' n'} & \simeq \frac{\ell' \left(n^2 + n +2\ell^2 +\ell -2 \right)}{\ell \left(n'^2 + n' +2\ell'^2 +\ell' -2 \right)}  \label{eq:F_WKB}\,, \\
	\T_{\ell n \ell' n'} & \simeq \frac{2n +1}{2n'+1} \label{eq:T_WKB}\,.
\end{align}
In figure~\ref{Fig:lmn-220}, we show these values with dashed gray gridlines and see that they are qualitatively consistent with the values of the ratios for $a=0$.

\begin{figure}
\centering
\includegraphics[width=\linewidth]{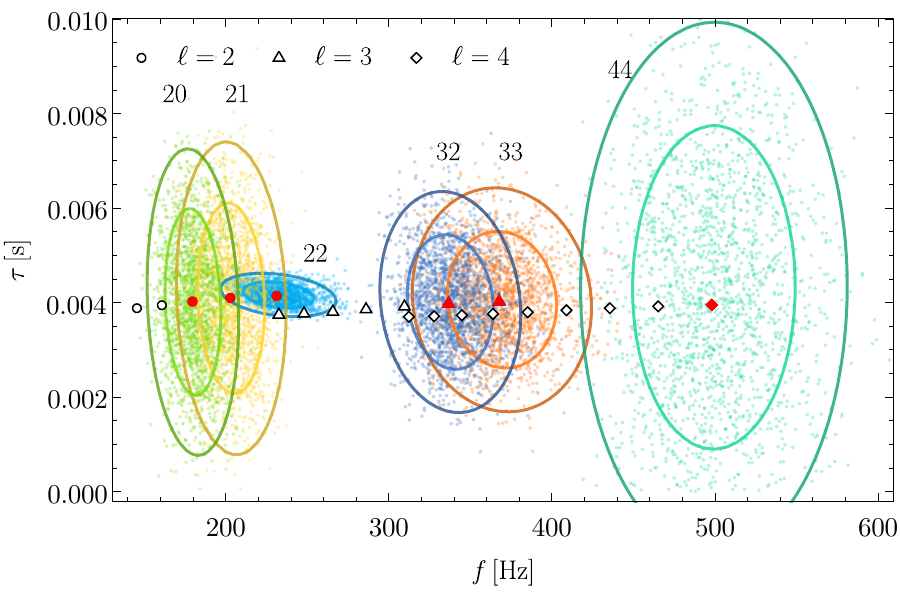}
\includegraphics[width=\linewidth]{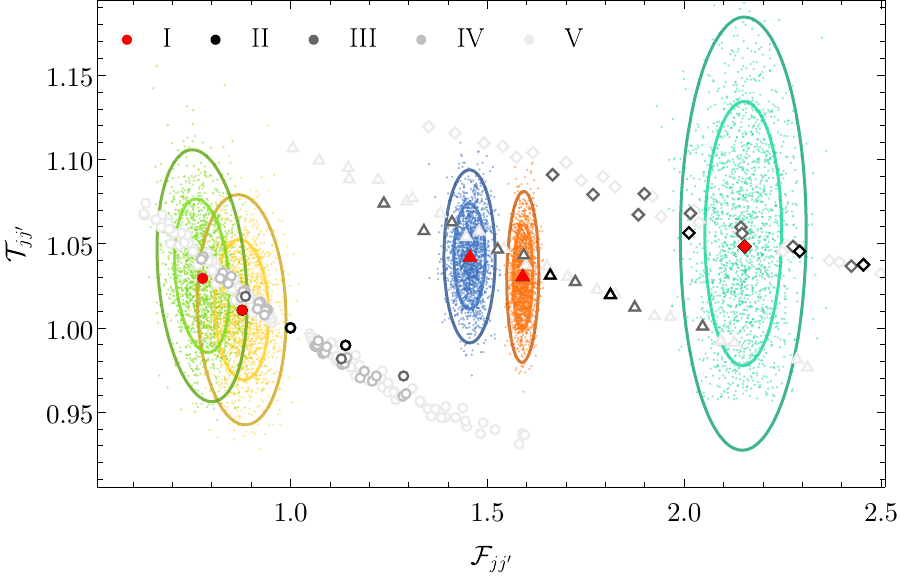}
\caption{Detectability forecast for QNM ratios of $(\ell,m)=(2,2),(2,1),(2,0),(3,3),(3,2),(4,4)$ modes for a GW150914-like binary with $q=2$ and ringdown SNR=$85$ as if detected by ET. The upper plot shows distributions for $f_{\ell m}$ and $\tau_{\ell m}$, while the lower plot the ratios $\F_{\ell m 2 2}$ and $\T_{\ell m 2 2}$. The light (dark) contours show the $1\sg$ ($2\sg$) confidence level of a bi-normal distribution of $2\times10^3$ randomly sampled points. Red dots show the injected values. In the top panel we show with black empty markers GR values of the other $m$ for same mass and spin of the injected values. In the lower panel for $\F_{j 22}$ and $\T_{j 22}$ contours and samples, the colours for each $j$ correspond to those of the upper panel. We also show the GR predictions of ratios between other mode combinations at the spin of the injection $a=0.62$, split in different classes of relevance as described in the text. \label{Fig:forecast}
}
\end{figure}

\noindent \textbf{\textit{Application to future detectors:}}~We now want to assess the distinguishability between modes with future detectors, for which we present a forecast measure on Einstein Telescope (ET)~\cite{Abac:2025saz}. We take the Fisher analysis mode detectability estimates from~\cite{Bhagwat:2023jwv}, where the authors assumed the measurement of one mode of the $(\ell,m) = (2,1),(3,3),(4,4)$ on top of the $(2,2)$\footnote{We drop the index $n=0$ as we do not consider overtones in this analysis} for a GW150914-like event with a ringdown SNR larger than $60$. The best case reported is that for mass ratio $q=2$, final mass $M_f = 70M_\odot $, final spin $a_f = 0.62 $ and $\mathrm{SNR}=85$. We propagate the estimated errors for $\F_{j j'}$ and $\T_{j j'}$ sampling randomly on $2\times10^3$ points, assuming $j = (2,1),(2,0),(3,3),(3,2),(4,4)$ and $j'=(2,2)$. For the $(2,0)$ and $(3,2)$ modes, which are not reported in~\cite{Bhagwat:2023jwv}, we assume a measurement with the same errors as the $(2,1)$ and $(3,3)$ modes, respectively. 

In figure~\ref{Fig:forecast}, we show the samples along with $1\sg$, $2\sg$ confidence levels for both frequency and damping times (upper panel) and their ratios (lower panel), plotted over the GR expectation at the injection value. We also present GR values that might be confused with those of the injection: in the upper panel we add the QNMs with same $\ell$ of the injections and all possible values of $m$, with black empty markers; in the lower panel, we report all possible mode combination ratios in this region, categorized by a grayscale ranking system, as follows (using $\ell=[2,4]$ throughout):
\begin{itemize}
	\item I:   GR injection;
	\item II:  $\ell' = 2$, $\ell -1 \leq m \leq \ell $, $1 \leq m'\leq 2$: the primary mode has the lowest angular momentum and both modes have the two highest combinations of azimuthal numbers;
	\item III: $\ell' = 2$, $m \geq 0$, $1 \leq m'\leq 2$: same as above, but with all possible combinations of non-negative azimuthal numbers;
	\item IV:  $\ell' = [3,4]$, $m \geq 0$, $m' \geq 0 $: same as above, including a primary mode with larger angular momentum;
	\item V:   $\ell' = [2,4]$: all possible combinations of $m$ and $m'$.
\end{itemize}
This ranking is based on the fact that negative-$m$ modes are highly suppressed in non precessing binaries and higher angular momentum modes are subdominant for comparable mass ratio systems. Hence, for the following discussion, we focus only on cases I-II-III.

Let us comment the results of the Fisher estimates. The upper panel shows degeneracy between the $(2,2)$ and $(2,1)$ and between the $(2,1)$ and $(2,0)$ modes, as well as between the $(3,3)$ and $(3,2)$ modes, plus the other modes not considered by the injections. Provided that in the analysis one selects the $(2,2)$ as the dominant (\textit{i.e.}, the mode whose amplitude is the largest), one can produce ratios and obtain the results of the lower panel. The degeneracy between the $(2,1)$ and the $(2,0)$ persists yet is more marginal, while the one between the $\ell=3$ modes is broken. It is worth noting that very few cases belonging to class I-II-III fall within the $2\sg$ contours of the injected modes:
\begin{itemize}
	\item for the $(2,1)$: the cases $(2,0)/(2,2)$, $(2,0)/(2,1)$;
	\item for the $(3,3)$: the cases $(3,2)/(2,1)$;
	\item for the $(3,2)$: the case  $(3,0)/(2,1)$;
	\item for the $(4,4)$: the cases $(4,3)/(2,2)$, $(4,m)/(2,1)$ with $m=[1,3]$, $(4,m)/(2,0)$ with $m=[0,1]$.
\end{itemize}

Key conclusions are: first, all the three cases exclude the detection of an overtone. Secondly, from this analysis we can clearly distinguish between diffent $\ell$ modes, with the only remaining degeneracy concerns the azimuthal number $m$.

\begin{figure*}
\centering
\includegraphics[width=\linewidth]{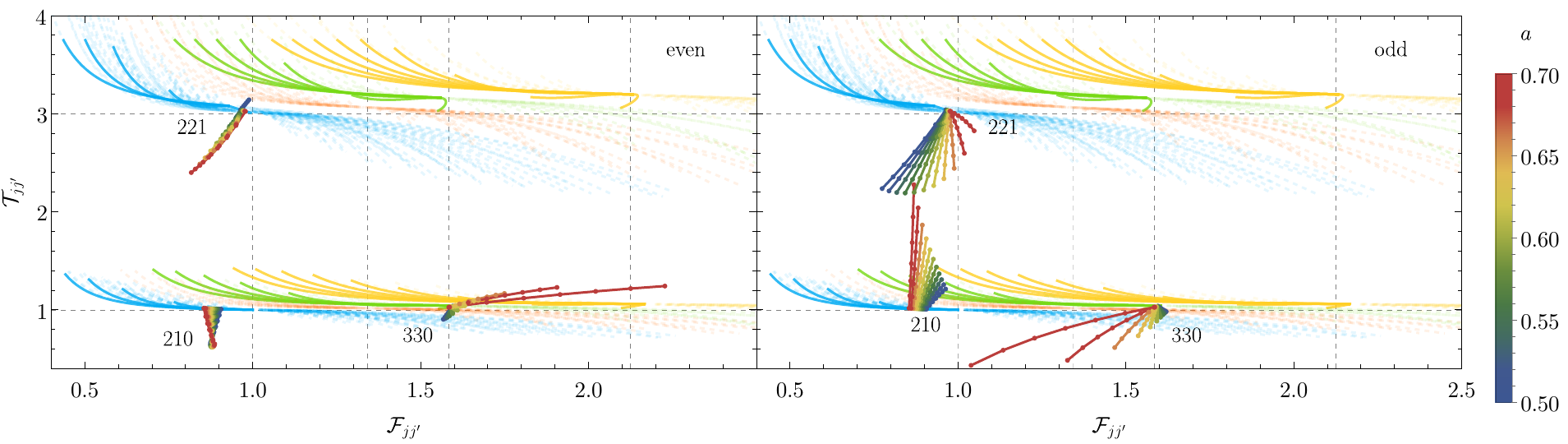}
\caption{Comparison of $\F_{j j'}$ versus $\T_{j j'}$ in GR and HDG. For GR, we plot the same mode combinations as in figure~\ref{Fig:lmn-220}, whereas for HDG we selected $j=[(2,2,1),(2,1,0),(3,3,0)]$, $j'=(2,2,0)$ for even/odd-parity perturbations displayed in the left/right panel. For each HDG set of frequencies, we span $a=[0.5,0.7]$ with steps of $\de a = 0.01$ corresponding to different colors (refer to the bar legend) and we span the coupling constant of the theory as $\la=[0,0.3]$ and each step in $\de\la=0.05$ is marked by a dot in each curve. \label{Fig:HDG}
}
\end{figure*}

\begin{figure}
\centering
\includegraphics[width=\linewidth]{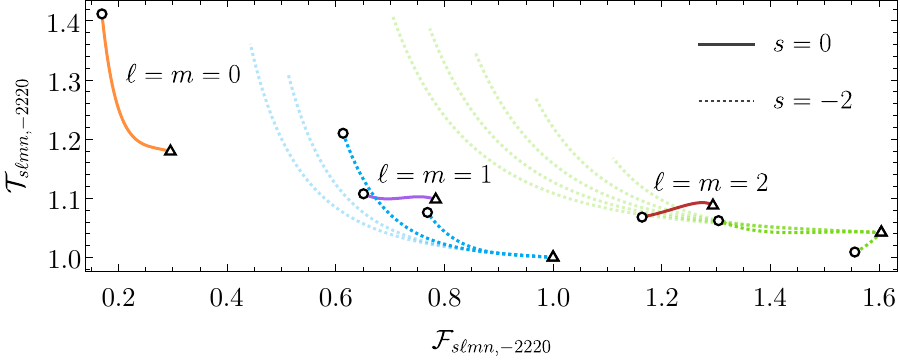}
\caption{Comparison of frequency and damping time ratios for scalar modes $s=0$ over the tensor $(-2,2,2,0)$ mode (solid lines). We also report combinations of $\ell=[2,3]$ modes over the $(-2,2,2,0)$ mode (dotted lines) with the same color scheme of figure~\ref{Fig:lmn-220}. Open triangles mark the ratios at $a=0$, circular ones at $a=0.9$.
\label{Fig:scalar_ratios}
}
\end{figure}

\noindent \textbf{\textit{Beyond-GR examples:}}~Lastly, we provide an example where the QNM predictions for a theory beyond GR yield ratios that are drastically different from the GR predictions. Specifically, we compare GR ratios against those in higher-derivative gravity (HDG), a theory of gravity containing higher-derivative curvature corrections to the action (see~\cite{Cano:2019ore} for more details). We focus on the third-order even-parity theory, corresponding to the coupling proportional to $\la_\mathrm{ev}$ in~\cite{Cano:2023tmv}, hereafter denoted simply $\la$. It is worth noting that the additional terms in the action break the isospectrality of the Kerr spectrum: even-parity and odd-parity modes become non-degenerate in this theory.
Using the linearized QNMs computed in~\cite{Cano:2023jbk,Cano:2023tmv,Cano:2024ezp}, figure~\ref{Fig:HDG} compares $\F_{j (2,2,0)}$ and $\T_{j (2,2,0)}$ for $j=[(2,2,1),(2,1,0),(3,3,0)]$ for different values of the spin and the coupling constant $\la$ overlaid on the GR combinations from figure~\ref{Fig:lmn-220}. The left/right panel refers to even/odd-parity modes, respectively.

For the HDG QNM ratios, we consider spins $a$ in $[0.5,0.7]$, conservatively chosen to represent astrophisically relevant BHs. The coupling constant $\la$ varies in $[0,0.3]$. Figure~\ref{Fig:HDG} shows that all mode combinations lie outside the range consistent with GR prediction. This deviation is particularly pronounced for the first overtone, which falls ouside the GR region for both even and odd-parity perturbations across nearly all parameter values. For the $(2,1,0)$ mode, the even-parity branches away immediately from the GR values, whereas odd-parity perturbations overlap the region spanned by $\ell/\ell'=3/2$ modes in GR. However, this degeneracy is broken at larger couplings. The $(3,3,0)$ does not overlap with the GR region only for large values of the spin and moderate values of the coupling constant.

Finally, it is well known that coupling between different fields with spin $s$ leads to spectrum contamination~\cite{Pani:2009wy,Yunes:2009hc,Pani:2011gy,Maselli:2015tta,Pierini:2021jxd,Pierini:2022eim,Wagle:2021tam,Srivastava:2021imr,Crescimbeni:2024sam,Lestingi:2025jyb}. Here, we report the case of a scalar field mode appearing in the gravitational field spectrum, assuming that the beyond-GR deviations to the spectrum are subdominant, and hence comparing $s=0$ QNMs against $s=-2$. In figure~\ref{Fig:scalar_ratios}, we show the ratios between scalar $\ell = [0,2]$, $m=\ell$ modes over the $(2,2,0)$ gravitational mode, compared with the ratios presented in figure~\ref{Fig:lmn-220}. It appears that the $\ell =0$ ratio lies in an empty region, while $\ell = [1,2]$ ratios partially overlap on the region spanned by ratios of gravitational modes, but with $\ell +1$. Hence, if the gravitational spectrum is contaminated sufficiently by the monopolar scalar mode, it would clearly appear from this analysis.

\noindent \textbf{\textit{Conclusion:}}~In this Letter, we offer a new possibility to test GR via ringdown analysis, based on a pure damped sinusoid fitting. Our method compares the ratios of the frequencies and damping times between at least two detected modes. The ratio automatically removes mass dependence, while we have demonstrated that the spin mildly affects the ratio of the real parts of the frequencies, at least for values of the spin which are relevant in binary BH mergers. We have shown that in GR, the ratio of the real part of any two modes is largely dominated by the angular momentum number $\ell$, while the ratio of the imaginary part is mostly determined by the overtone number $n$. We provided an euristic explanation of this behaviour from a WKB eikonal perspective. We showed how these universal relations hold in GR but are not respected in many beyond GR examples or if additional fields couple non-minimally to gravity. This test provides insightful implications in GR too, as it allows to assess or exclude the presence of some mode combinations where normally damped sinusoid analysis marginally been used, as for detection of overtones or quadratic modes. We provided examples in which standard spectroscopy fails while QNM ratios allow a better distinction of modes. With a golden event detected by ET with the most agnostic setup for ringdown, one can distinguish among different $n$ and $\ell$ having completely removed any degeneracy with the value of the final mass.

This analysis offers direct applications to current spectroscopy data analysis, as the damped sinusoid ringdown model is already part of the LIGO-Virgo-Kagra test of GR routine. 
We remark that to have full mode identification in an agnostic setup, one should also combine results of QNM ratios with the amplitude-phase consistency test proposed in~\cite{Forteza:2022tgq}. 
The generality of this test allows to extend the damped sinusoid analysis even to the beginning of the postmerger to identify the evolution of ringdown signal from the peak to the linear regime. In this respect, it will be crucial to have a description of the prompt response, which can then be incorporated in the test (for recent progress, see~\cite{DeAmicis:2025xuh,Oshita:2025qmn,Kuntz:2025gdq,Arnaudo:2025uos}).
Finally, in this paper we presented the test with the assumption of a 2-modes damped sinusoid model, so it would be crucial to understand the feasibility and advantages of a multimodal QNM ratios analysis.

\acknowledgments

I am indebted to Giada Caneva Santoro, Gregorio Carullo, Vasco Gennari, Joachim Pomper, Thomas Sotiriou and Sebastian V\"olkel for fruitful discussions.
I acknowledge funding from the FCT grant agreement 2023.06263.CEECIND/CP2830/CT0004, and the support to the Center for Astrophysics and Gravitation (CENTRA/IST/ULisboa) through FCT grant No.~UID/99/2025.

\bibliography{literature}

\end{document}